# Electrical spin injection into graphene from a topological insulator in a van der Waals heterostructure

Jifa Tian[1,2], Ting-Fung Chung[1,2], Ireneusz Miotkowski[1], and Yong P. Chen[1,2,3,4,*]

[1]Department of Physics and Astronomy, Purdue University, West Lafayette, IN 47907, USA

[2]Birck Nanotechnology Center, Purdue University, West Lafayette, IN 47907, USA

[3]School of Electrical and Computer Engineering, Purdue University, West Lafayette, IN 47907, USA

[4]Purdue Quantum Center, Purdue University, West Lafayette, IN 47907, USA

*Email: yongchen@purdue.edu

**Abstract**

All-electrical (magnetic-material-free) spin injection is one of the outstanding goals in spintronics. Topological insulators (TIs) have been recognized as a promising electrically controlled spin source thanks to the strong spin-orbit coupling and in particular, the spin-momentum locked topological surface states (TSS) supporting helically spin polarized currents. Many TI materials such as Bi-based chalcogenides are also layered 2D materials and can be incorporated into van der Waals (vdW) coupled heterostructures, opening the possibility of the utilization of TIs for electrical spin injection into other 2D materials. Here, we demonstrate electrical injection of helically spin-polarized current into graphene through a 3D TI in a mechanically stacked heterostructure between $Bi_2Te_2Se$ (a TI) and chemical vapor deposition (CVD)-grown graphene, using the spin potentiometric measurement. When a dc current is flowing from the TI to graphene, we detect a striking step-like voltage change (spin signal) in both the TI and graphene using ferromagnetic (FM) probes. The sign of the spin signal can be reversed by reversing the direction of the dc bias current, and the corresponding amplitude of the spin signal increases linearly with the bias current, indicative of a current-induced helical spin polarization in both TI and graphene. In contrast, the graphene itself exhibits usual nonlocal spin valve signal when the spins are injected using an FM electrode. We discuss possible origins of the helical spin polarization injected into the graphene in our TI/graphene heterostructure that may include TSS as well as the spin-orbit coupled Rashba states. Our findings show electrical injection of a spin helical current into graphene through a TI and demonstrate TIs as potential spin sources for future spintronic devices wherein spin manipulation is achieved electrically.

**Introduction**

The advent of van der Waals (vdW) heterostructures based on atomically thin crystals, such as graphene and other 2D materials [1,2], has generated strong scientific interests with potentials for novel device technologies [3]. For example, using the powerful idea of band-structure engineering, one can combine several different 2D materials to create a designer potential profile for electrons. This has led to many recent device demonstrations such as tunneling diodes with negative differential resistances [4], tunneling transistors [5], photovoltaic devices [6,7], and so on. However, such vdW heterostructures have so far rarely been employed in spintronics, although some of the 2D materials themselves have shown great potentials in applications for future spintronic devices. For example, graphene, an atomic layer of carbon atoms with weak spin-orbit coupling (SOC), has been shown as one of the prominent candidates with long spin diffusion lengths of up to several μm's at room temperature [8]. This makes graphene a promising conduction channel for carrying spin-related information. Another example is topological insulators (TIs), which host non-trivial topological surface states (TSS) where charge carriers are massless Dirac fermions with their spins locked in-plane and perpendicular to the momentum (Fig. 1a). [9-15] Real TI materials may also contain other conducting states, such as a 2D electron gas (2DEG) with Rashba SOC induced by the bending of bulk conduction band near the surface, that may also give current-induced spin polarization [16-18]. Recently, current-induced helical spin-polarization in 3D TIs has been demonstrated in various Bi-based TIs [19-26] such as bulk metallic $Bi_2Se_3$ [19,21,23], bulk insulating $Bi_2Te_2Se$ (BTS221) [20] and $Bi_{1.5}Sb_{0.5}Te_{1.7}Se_{1.3}$ [25], using ferromagnetic (FM) contacts as spin detectors in different electrical configurations. Most importantly, the amplitude and direction of a dc charge current govern the amplitude and direction of the resulting spin polarization [19-22], making TI an appealing source for generating electrical spin polarization without any FM materials.

Previous works on graphene spintronics mostly used FM contacts to inject polarized spins. Combining graphene and 3D TIs into heterostructures (Fig. 1a) opens a new possibility for spins to be injected electrically (instead of using FM) into graphene through a charge current applied between the TI and graphene. However, spin injection from a 3D TI into a nearby 2D material, while theoretically predicted [27,28], has only been experimentally demonstrated very recently in a TI/graphene (TI/G) heterostructure fabricated by growing a TI flake using chemical vapor deposition (CVD) on top of an exfoliated graphene [29] flake and measured using a non-local spin detection method. Here, we demonstrate such spin injection in a high quality TI/graphene heterostructure fabricated by the mechanical stacking method (transferring TI on graphene), and using a different measurement technique (spin potentiometry) to directly detect the spin polarization of the injected helical spin current. We demonstrate that a spin helical current can be successfully injected into graphene through a bulk insulating BTS221 (TI) thin flake in a TI/G heterostructure (Fig. 1b), using 3D TI as a spin generator and graphene as a spin transport channel. We applied a dc bias current across the TI/G junction and utilized Co tunneling contacts on the graphene and TI surfaces for detecting the channel spin polarization (so called the spin potentiometric measurements). Similar to that in TI alone, we observed a step-like voltage change (spin signal) in both the TI and graphene sections of the TI/G heterostructure. We find that the sign (and amplitude) of the spin signal is determined by the direction (and amplitude) of the dc bias current. While our results make a direct demonstration of electrical injection of a helical spin-polarized current into graphene through a 3D TI, the exact origins of the spin helical current remain to be better understood. We discuss various possibilities, such as the TSS on the TI top surface, the Rashba 2DEG states formed at the TI bottom surface, and graphene itself with enhanced SOC at the TI/G interface.

**Experimental**

**Sample fabrication.** TI and graphene (TI/G) heterostructures (Fig. 1b) were fabricated by mechanical stacking techniques (Fig. S1). The graphene was grown on a Cu foil by atmospheric pressure CVD method [30]. After the Cu foil was completely etched away, the CVD graphene was transferred onto a heavily doped Si/SiO$_2$ substrate [30]. The TI thin flakes of typical thickness of 20 nm were exfoliated from an n-type bulk BTS221 crystal (grown by Bridgman method) [31] and placed on a polyvinyl alcohol and polymethyl methacrylate (PVA/PMMA) film [32]. Then, the PVA/PMMA thin film was transferred on top of the CVD-grown graphene with the BTS221 thin flake (bottom surface) directly contacting the graphene surface (Fig. S1a). Later, the PVA/PMMA film was dissolved in acetone and rinsed in isopropanol. The rectangle shapes of the BTS221 and graphene in the heterostructure (Figs. 1b,S1b) were defined by e-beam lithography and etched by Ar and O$_2$ plasma (Fig. S1b), respectively. Ferromagnetic (FM, 50 nm thick Co) and non-magnetic (70 nm thick Au) contacts were fabricated by two rounds of e-beam lithography and e-beam evaporation. To make FM tunneling contacts, a 0.6 nm thick Al film was deposited and fully oxidized after exposing to air before depositing Co contacts. The representative device structure is shown in Fig. 1b.

**Electrical measurements.** TI/G heterostructure devices were measured in a variable temperature insert with a based temperature of T = 1.6 K. In order to check the quality of the BTS221 flake in the TI/G heterostructure, a spin potentiometric measurement [18,20] was performed on the BTS221 part, where a dc bias current was applied between two Au ("1" and "6") contacts and a voltage (spin signal) was measured between Co ("4") and Au ("5") contacts. To confirm good spin transport properties of graphene in the TI/G heterostructure, we performed the typical non-local spin valve measurements on graphene, where a bias current was applied between Co ("2") and Au ("1") contacts and a non-local voltage (spin signal) was measured between Co ("3") and Au ("5") contacts. After these two sets of experiments, we performed the spin potentiometric measurements on the BTS221/graphene heterostructure (Fig. 1b), where a dc bias current was applied between two Au ("1") and ("6") contacts and a voltage was measured between Co ("3") and Au ("5") contacts.

**Results and Discussion**

Representative results from the spin potentiometric measurements (Fig. 2a) performed on the TI part (BTS221) at different bias currents are shown in Figs. 2b-d. At a positive dc bias current I = 20 µA, we observe a hysteretic step-like voltage change (spin signal), showing a low voltage state transitioning to a high voltage state as the magnetization $M$ of the Co contact changes from $+M$ to $-M$, as shown in Fig. 2b (directions of +B field and +I are labeled in Fig. 2a). Reversing the direction of the dc bias current (I= -20 µA) reverses the trend of the transition between the high and low voltage states as shown in Fig. 2c. From the electronic chemical potential analysis [20], we know that the high and low voltage states are determined by the relative orientation between the channel spin polarization ($s$, induced by a dc bias current) and the magnetization ($M$) direction of the Co spin detector. The results measured in BTS221 flake of the TI/G heterostructure are consistent with our previous studies of the spin-momentum-locked TSS in 3D TI thin flakes [20], whereas a high (low) voltage state is always observed when the channel spin polarization $s$ is parallel (antiparallel) to the direction of $M$ (Figs. 2b,c). To further confirm our conclusions, we systematically studied the current dependence of the spin signal by varying the dc bias current from 0 to ± 300 µA at T=1.6 K. The representative results are shown in Figs. 2d,S2, where we can see that the spin signal, δV=V(+$M$)-V(-$M$), increases approximately linearly with increasing dc bias current, consistent with the expectations of the spin-momentum-locking of TSS. Within the proposed model of spin potentiometry [18], the spin signal δV is given by $\delta V = (V(+M) - V(-M)) = IR_B P_{FM} (\boldsymbol{p} \cdot \boldsymbol{M_u})$. Here, I is the bias current, $1/R_B$ is the "ballistic conductance" of the channel given by $e^2/h$ times the number of conducting modes ($\sim k_F W/\pi$ for each Fermi surface, where $k_F$ is the Fermi wave number, $W$ is the width of channel). $\boldsymbol{M_u}$ is the unit vector along the positive detector magnetization direction. We can estimate the spin polarization $p$ of the surface current [18,20] to be $|p| \approx 0.5$ from the slope of δV vs. bias current $I$ in Fig. 2d based on the corresponding channel width W = 7 µm, $k_F \approx 0.1$ Å$^{-1}$ (estimated from ARPES measurements on our BTS221 bulk crystals [27] as well as Hall measurements in flakes similar to those used here). For this order of magnitude estimate, we have assumed $P_{FM}(Co) \approx 0.4$ [19] and also that about half of the total current flows through the top surface channel (which contacts our Co spin detectors, and the other half of total current $I$ flows through the TI bottom surface and the CVD graphene, noting the TI bottom surface is in direct contact with graphene). The above results confirm the existence of the current-induced spin polarization in the BTS221 flake of the TI/G heterostructure.

To check the spin transport properties of the graphene part in the TI/G heterostructure, we performed the typical non-local spin valve measurements [8] on graphene (Fig. 3a). The representative results are shown in Figs. 3b. We can see that, when the Co ("2" &"3") contacts are magnetized to the down direction (indicated by green arrows in Fig. 3b), the measured non-local voltage $V_{NL}$ (red curve) is relatively constant at ~ 10 µV, for example in the red-colored trace (upward B sweeps) at negative B fields. However, as the magnetization of one Co contact ("2") is reversed and points to the up direction at ~ 15 mT (whereas the magnetization of the other Co contact ("3") still points to the down direction), $V_{NL}$ abruptly decreases to ~ -7 µV, a low voltage state, as shown in Fig. 3b. The $V_{NL}$ remains at this low voltage state until the magnetization of Co ("3") contact is also reversed to the up direction at B ~ 25 mT. When B is larger than 25 mT, $V_{NL}$ changes back to the high voltage state of ~ 10 µV, whereas now both the Co ("2" & "3") contacts are magnetized to the up direction. Reversing the magnetic field sweep direction, the switching from high to low and back to high $V_{NL}$ (black curve) can be observed again, but now appearing at the negative magnetic fields near the coercive fields of the two Co ("2" & "3") contacts. We also applied a negative dc bias current of -5 µA and performed the similar non-local spin valve measurements on the sample. The corresponding results are shown in Fig. 3c, where the measured non-local voltage $V_{NL}$ vs. magnetic field $B$ is similar to that of I = 5 µA (Fig. 3b) but multiplied by a negative sign due to the negative dc bias current. If we convert $V_{NL}$ to the non-local resistance, $R_{NL} = V_{NL}/I$, the results measured at positive and negative currents give the similar $R_{NL}$ vs. B curves (Fig. S3). Furthermore, we also performed the similar non-local spin valve measurements on the graphene part by switching the current and voltage connections, such that the dc current was applied between Co ("3") and Au ("5") contacts and non-local voltage was measured between Co ("2") and Au ("1") contacts. Similar non-local spin valve results were observed, as shown in Fig. S4. We found that the non-local spin signal can be observed in our CVD graphene up to room temperature. In our measurements, the gap between the two Co ("2" & "3") contacts shown in Fig. 3a is ~ 600 nm, suggesting that the spin diffusion length in our CVD-grown graphene significantly exceeds 600 nm. Our non-local spin valve results demonstrate the good quality of CVD graphene for the spin transport in the TI/G heterostructure.

After the two sets of experiments on the TI and graphene parts of the TI/G heterostructure, we performed the spin potentiometric measurements (Fig. 4a) across the TI and graphene junction. The representative results measured at I = ± 200 µA are shown in Figs. 4b,c, where we can observe striking hysteretic step-like voltage changes as the in-plane magnetic field is swept between ±0.1 T. In Fig. 4b, the measured voltage switches from a high voltage state to a low voltage state as the magnetization *M* in the Co ("3") contact is switched from –*M* to +*M*. Most importantly, reversing the current direction reverses the trend of the voltage change as shown in Fig. 4c, whereas now the low voltage state transitions to the high voltage state as the magnetization *M* of Co ("3") contact is switched from -*M* to +*M*. Such a step-like voltage signal indicates electrical detection of a current-induced helical spin polarization. [19-26] Furthermore, we can determine the direction (labeled by the red arrows in Fig. 4bc) of channel spin polarization from the sign of the step-like spin signal, which reverses upon current reversal. Such a step-like spin signal is not expected in graphene alone due to its weak SOC. We also rule out the local Hall effect induced by fringe fields of the Co contacts [33] as a cause of the step-like spin signal as it has not been observed in the CVD graphene alone using the similar spin potentiometry. The spin potentiometric measurement in our TI/G heterostructure directly demonstrates that a helical spin-polarized current has been successfully injected into graphene through the 3D TI (BTS221) thin flake.

Figure 4d shows the spin signal $\delta V$ as a function of the dc bias current *I*, where we see that $\delta V$ approximately linearly increases with *I*. Such a linearly dependent spin signal in the TI/G heterostructure is analogous with our previous results measured in the BTS221 flake alone as well as other 3D TI materials [19-21]. We also studied the temperature dependence of the spin signals measured at I= ±200 µA from 1.6 K to 60 K, with representative results shown in Figs. 4e,S5. We can see that the spin signals $|\delta V|$ for both positive and negative bias currents are only weakly dependent on temperature in the measurement range. As the temperature further increases, the spin signal $\delta V$ is expected to decrease due to the increased conduction from bulk states.

The directions of the injected spin polarization *s* in the graphene part of the TI/G heterostructure, as inferred from the sign of $\delta V$ and the direction of the magnetization *M* in the Co spin detector for positive and negative dc currents (I = ±200 µA), are shown in Figs. 4b,c respectively. Interestingly, the spin polarization direction for a given current direction, or the sign of the spin helicity, is observed to be the *same* as that measured in the BTS221 part (shown in Figs. 2b,c and consistent with the expected spin helicity of TSS on the *top* surface of the TI [20]). Such an observation may suggest that the spin helical electrons from the TSS on the top surface of the TI flake are injected into the graphene, since the TSS on the bottom surface of the TI features an opposite spin helicity [11,12]. However, it is important to consider the role of the bottom surface of the TI flake where graphene makes a direct contact in our structure. The work functions of BTS221 and single layer graphene are $\phi_{BTS221} = 5.23$ eV [34] and $\phi_G = 4.57$ eV [35], respectively. Therefore, when graphene is in contact with the TI flake (Fig. S6), the aligning [36] of their chemical potentials causes a downward bending of the TI bands at the interface (Fig. S6). Since the amount of the downward bending, given by $\phi_{BTS221} - \phi_G = 0.66$ eV, exceeds the bulk bandgap (~0.3 eV [31]) of the BTS211, a 2DEG (derived from the TI bulk conduction band) would form near the interface (bottom surface of the TI in our structure). Such a band-bending induced 2DEG [16,17] has been shown to possess strong Rashba SOC, and its outer Fermi surface (which typically dominates the transport [18]) has the opposite spin helicity [16,17] compared to that of TSS on the same surface. This suggests that the Rashba 2DEG at the TI *bottom* surface, whose dominant spin helicity is the same as that of the TSS on the TI *top* surface, provides another possible source of the observed spin-helical current injected from TI into graphene in our structure. The possibility of a combined contributions from both the top and bottom surfaces of the TI is also supported by a lager slope (~ 1.7 V/A) of the spin signal δV vs *I* measured in the graphene with current injected from the TI (Fig. 4d) than that (~1.1V/A) extracted from the spin signal measured on the TI top surface alone (Fig. 2d). In addition, it is also possible for graphene to acquire enhanced SOC itself due to breaking the structural inversion symmetry or exchange/proximity coupling to TI at the TI/G interface that might enable a spin helical transport in the portion of graphene that is in direct contact with TI. Further studies are required to examine these different possible mechanisms and to more quantitatively extract the induced spin polarization of the spin-helical current injected from TI to graphene.

Recently, a current-induced spin polarization was reported in a TI/G heterostructure [29] fabricated by growing a TI thin film on top of an exfoliated graphene flake using the CVD technique. The corresponding spin signal is detected using a non-local method. The spin signal in graphene part injected from the TI flake was observed up to ~ 10 K. However, the spin helical transport in the TI part and the origins of the injected spin in graphene were not fully characterized [29]. In our work, we fabricated the high-quality TI/G heterostructures using the mechanical stacking method. The spin transport properties were studied in both the TI and graphene parts of the TI/G heterostructure. We find that the spin helical current injected from TI into graphene shows weak temperature dependence in our measured temperature range (up to ~ 60K) and the direction of the detected spin helical polarization on the graphene part is the same as that of TSS on the TI top surface. We suggest that the helically spin polarized electrons injected from TI to graphene could arise from both the TSS on the TI top surface and band bending induced Rashba 2DEG at the bottom surface. Both these two possible sources possess the same spin helicity as the one observed in our spin potentiometry measurements and can contribute together to the injected spin polarization. We also point out yet another possibility, to be clarified in future studies, that graphene itself in contact with the TI may acquire enhanced SOC that may also give rise to a current induced spin polarization.

**Conclusions**

In summary, we have fabricated a TI/graphene vdW heterostructure and demonstrated an electrical method to inject a spin helical current into CVD-grown graphene in the heterostructure through a 3D TI thin flake. The amplitude and direction of the injected helical spin polarization into graphene are determined by the charge current applied across the TI/G heterostructure. Our work gives a demonstration of utilizing a TI --- involving possibly both its TSS and Rashba 2DEG as spin sources --- to inject a spin helical current into graphene in our TI/G heterostructure, which may be applicable also to other 2D materials. Our work suggests that TI/2D material vdW heterostructures have the great potential for applications in future all-electric spintronic devices.

**Acknowledgements**

We acknowledge support for various stages of this work by a joint seed grant from the Birck Nanotechnology Center at Purdue and the Midwestern Institute for Nanoelectronics Discovery (MIND) of the Nanoelectronics Research Initiative (NRI), and by DARPA MESO program (Grant N66001-11-1-4107).

**Figure Captions**

**Figure 1 Topological insulator (TI) and graphene (G) heterostructure.** (a) Schematic diagram of the TI/graphene (TI/G) heterostructure, depicting the spin-momentum-locked Dirac cone of topological surface states (TSS) of TI and one of the two Dirac cones of graphene (spin degenerate with weak spin-orbit coupling). (b) A representative TI (exfoliated $Bi_2Te_2Se$, BTS221) and graphene (CVD-grown) heterostructure device used in our measurements. The ferromagnetic electrodes are Co, and the non-magnetic electrodes are Au with corresponding electrode numbers labeled.

**Figure 2 Spin potentiometric measurements on the TI (BTS221) part of the TI/G heterostructure device.** (a) Schematic diagram of a TI/G heterostructure device with corresponding 4-terminal electrical connections. In the measurement, a dc bias current (I) is applied between the two outmost Au electrodes ("1" and "6"), where the arrow represents the direction of +I. The voltage $V=V_4-V_5$ (for spin potentiometry) is measured between the Co electrode ("4") and the Au electrode ("5"). (b) The representative spin signal is detected as a step-like voltage change upon reversing the magnetization (M) of the Co electrode by sweeping an in-plane magnetic field (B) along its easy axis. The measurement is taken at I= 20 µA. The two traces in this panel (as well as in other figures showing two magnetic field dependent voltage traces) represent two opposite magnetic field sweep directions. (c) The voltage (spin potential) as a function of the in-plane magnetic field measured at I=-20 µA, showing a reversed sign of the spin signal. The spin signal $\delta V=V(+M)-V(-M)$ is indicated by the blue arrow. The directions of bias current (I), channel spin polarization (*s*) under the Co electrode and the magnetization (*M*) of the Co electrode are also labeled with corresponding arrows in different colors. (d) The current (I) dependence of the spin signal $\delta V$, showing an approximately linear increase with the increasing bias current. All the measurements are performed at T=1.6 K.

**Figure 3 Non-local spin valve measurements on the graphene part of the TI/G heterostructure device.** (a) Schematic diagram showing the electrical connections used for spin injection and detection in non-local measurements. A dc bias current is applied between the Co ("2") and Au ("1") electrodes, and a non-local voltage ($V_{NL}=V_3-V_5$) is measured between the Co ("3") and Au ("5") electrode. (b,c) The non-local voltage $V_{NL}$ as a function of in-plane magnetic field measured at I=5 µA (b) and I=-5 µA (c). The green arrows represent the directions of magnetization in the two Co ("3" & "2") contacts. All the measurements are performed at T=1.6 K.

**Figure 4 Electrical detection of spin helical current in graphene injected from TI (BTS221)**. (a) Schematic diagram of the spin potentiometric measurement with a 4-terminal connection, showing that the current induced helical spin polarization in TI is injected into graphene. A magnetic Co ("3") electrode on the graphene serves as the spin detector (the measured $V=V_3-V_5$). (b) The spin signal measured at I = 200 µA, showing a step-like voltage change upon reversing the magnetization direction of the Co electrode. (c) Reversing the bias current (I= -200 µA) reverses the trend of the hysteretic step-like voltage change, giving the direct evidence of injecting a spin helical current into graphene through the BTS221 flake. (d) Current dependence of the spin signal, exhibiting an approximately linear increase with the increasing dc bias current. (e) Temperature dependence of the spin signal measured at I=±200 µA, from 1.6K to 60 K. The measured spin signal is weakly dependent on temperature.

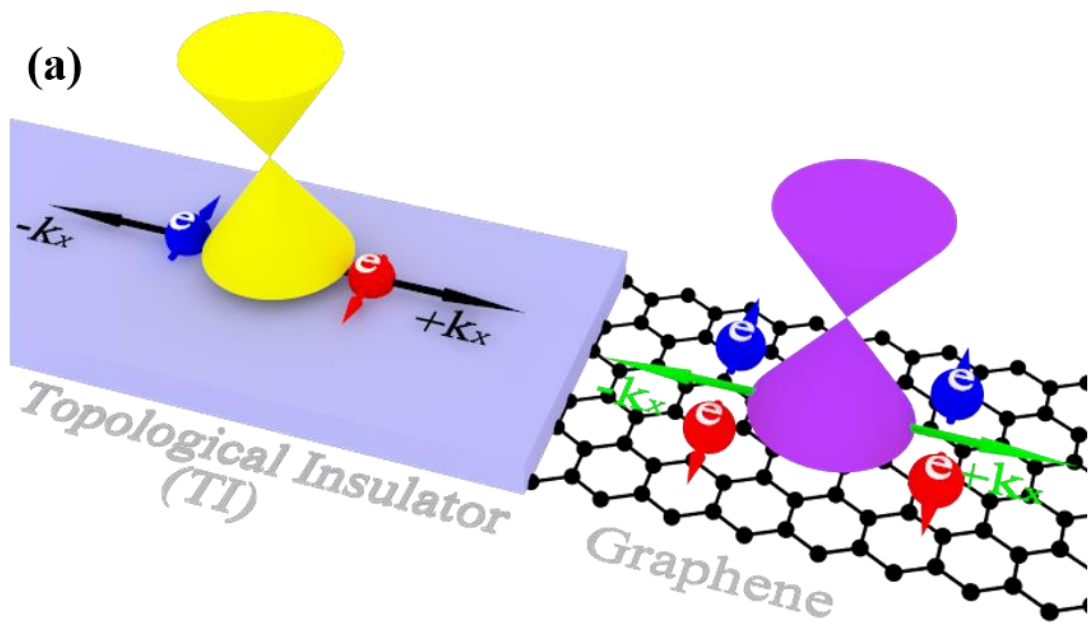
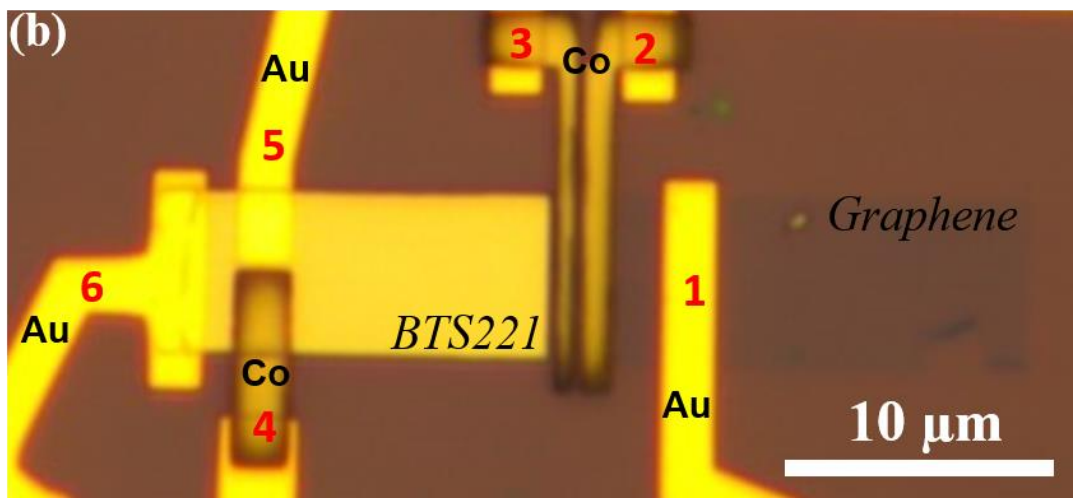

Figure 1 by Tian et al.

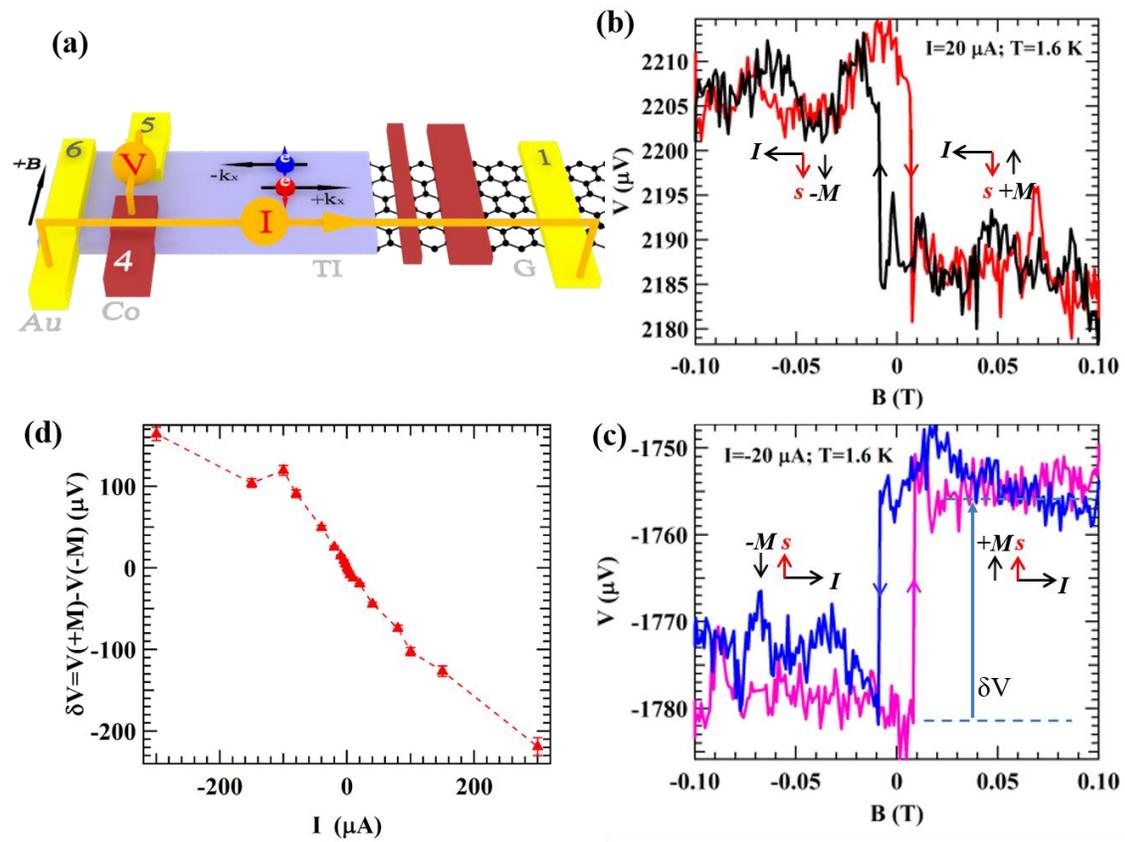

Figure 2 by Tian et al.

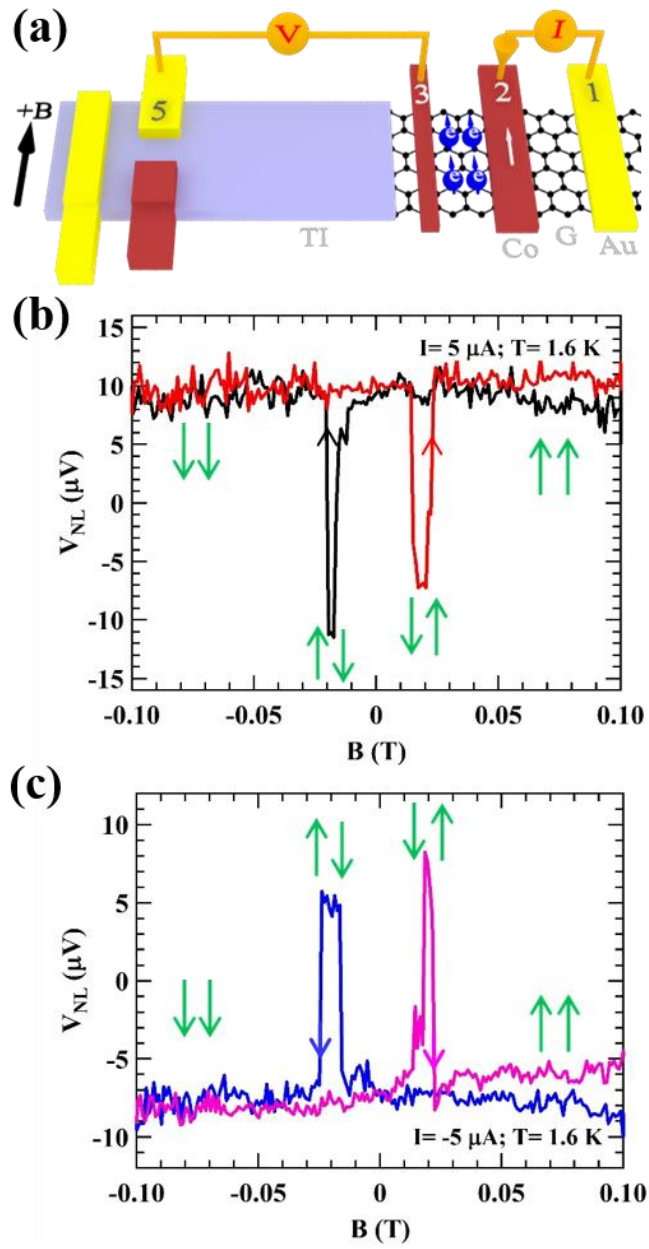

Figure 3 by Tian et al.

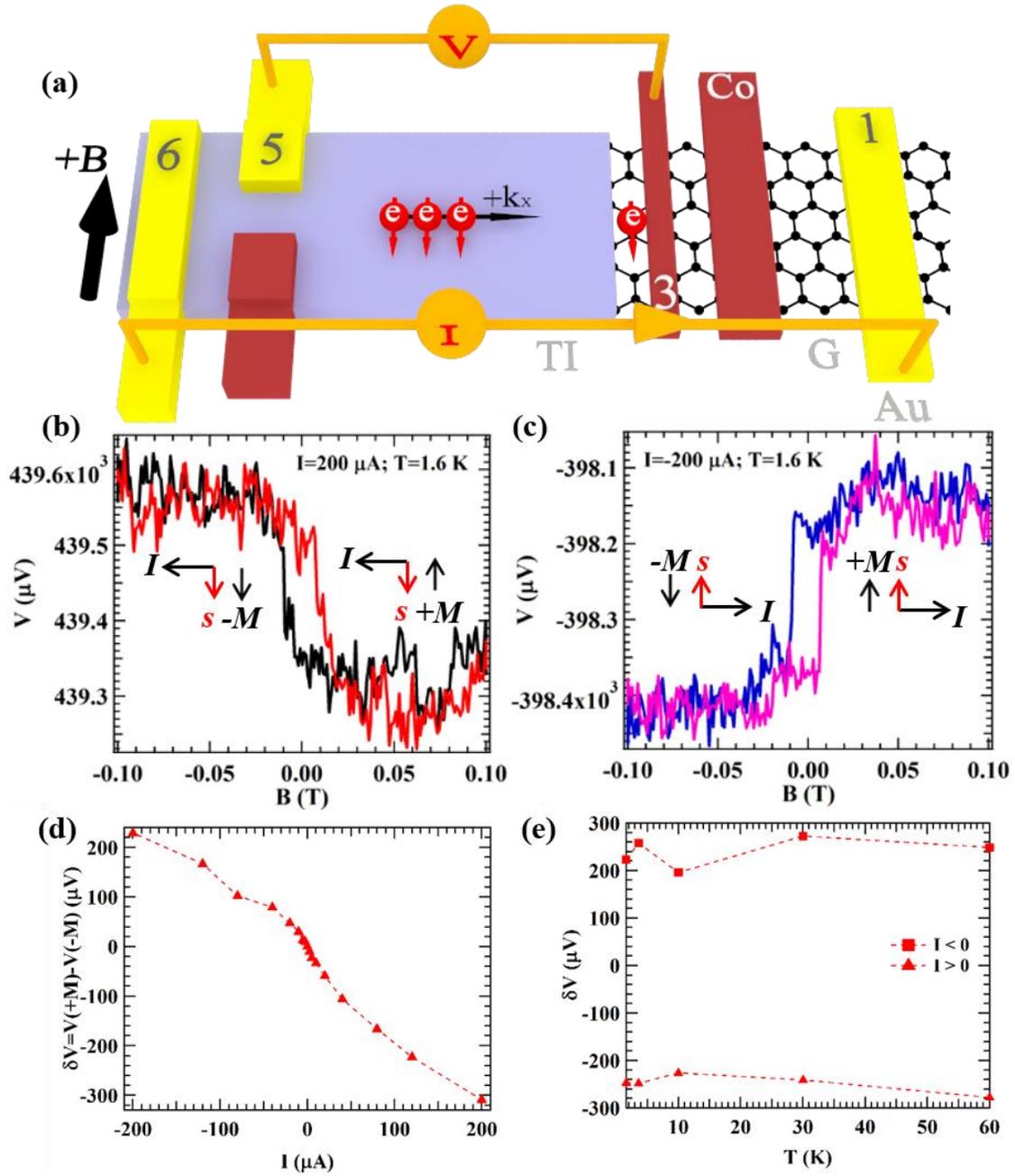

Figure 4 by Tian et al.

**Electrical spin injection into graphene from a topological insulator in a van der Waals heterostructure**

Jifa Tian[1,2], Ting-Fung Chung[1,2], Ireneusz Miotkowski[1], and Yong P. Chen[1,2,3,4,*]

[1]Department of Physics and Astronomy, Purdue University, West Lafayette, IN 47907, USA

[2]Birck Nanotechnology Center, Purdue University, West Lafayette, IN 47907, USA

[3]School of Electrical and Computer Engineering, Purdue University, West Lafayette, IN 47907, USA

[4]Purdue Quantum Center, Purdue University, West Lafayette, IN 47907, USA

*Email: yongchen@purdue.edu

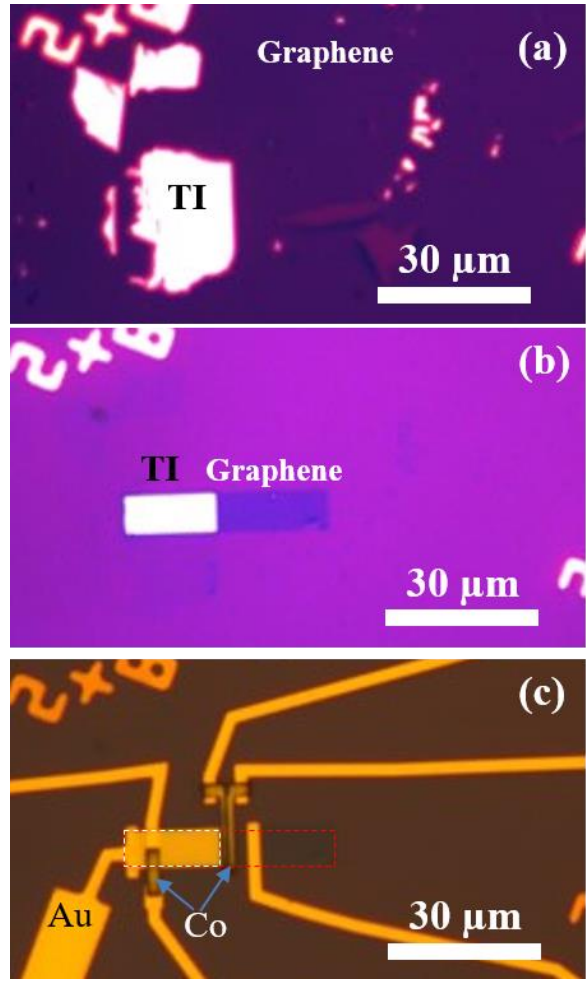

**Figure S1. Fabrication of topological insulator (TI) and graphene (TI/G) heterostructures.** (a) BTS221 thin flakes transferred on top of CVD-grown graphene on the SiO$_2$/Si substrate, where graphene covers the entire surface of SiO$_2$. (b) The BTS221 flake and graphene are further patterned into a rectangular shape (Fig. 1b) by e-beam lithography and plasma etching. The BTS221 flake was etched by Ar plasma and graphene was etched by O$_2$ plasma, consecutively. (c) The representative device image of the TI/G heterostructure after fabricating electrodes. The dashed white (red) rectangle represents the location of TI flake (graphene), respectively.

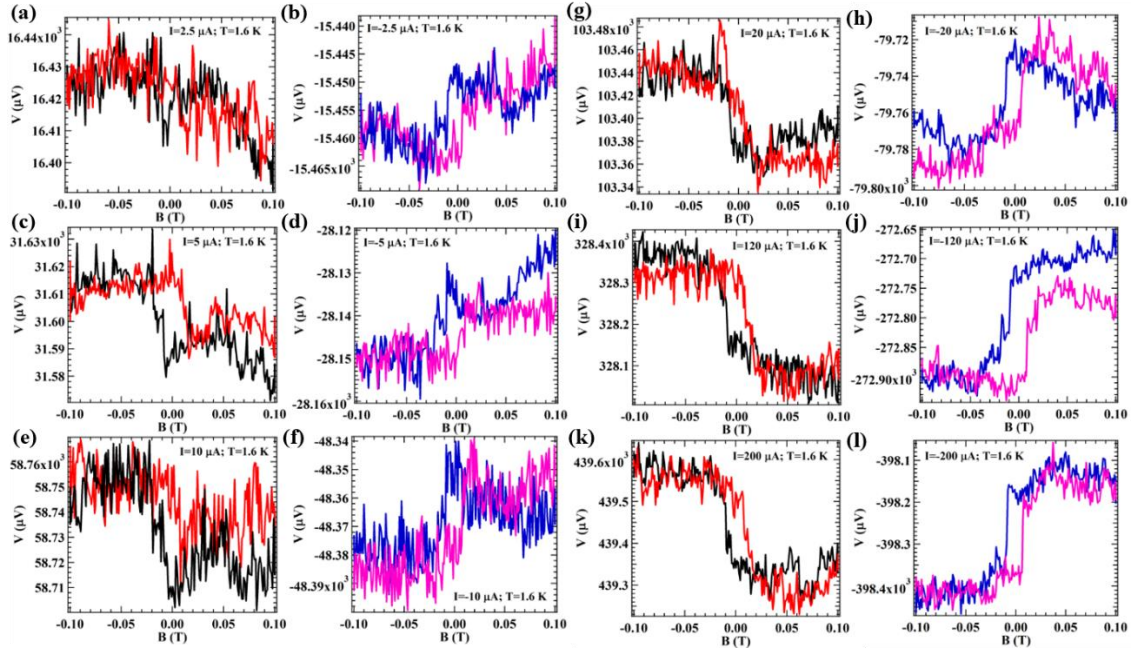

**Figure S2. Current dependence of the spin signal measured in the TI (BTS221) part of the TI/G heterostructure.** (a-I) The measured voltage (part of the datasets of Fig. 2d) as a function of the in-plane magnetic field measured at different bias currents of 2.5μA (a), -2.5 μA (b), 5 μA (c), -5 μA (d), 10 μA (e), -10 μA (f), 20 μA(g), -20 μA (h), 120 μA (i), -120 μA (j), 200 μA (k), and -200 μA (l), respectively. All the measurements are performed at T=1.6 K.

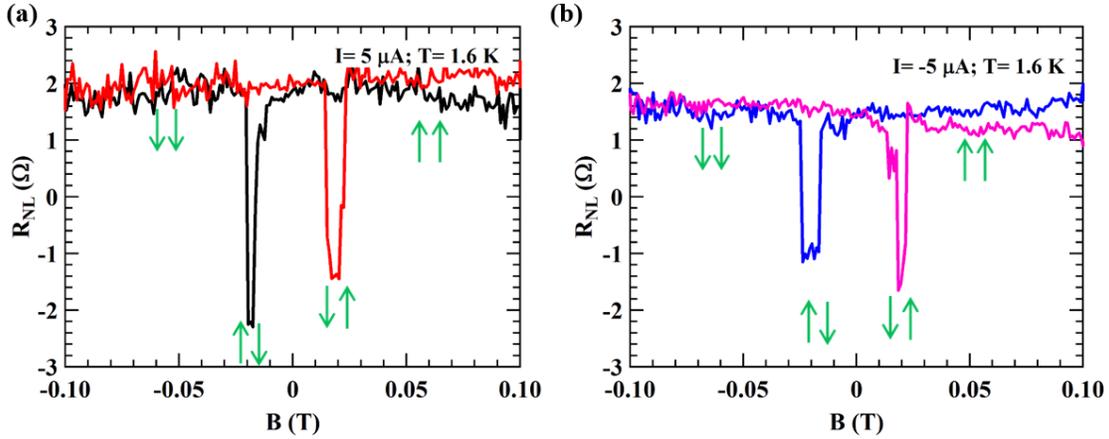

**Figure S3 Non-local spin valve (non-local resistance, $R_{NL}$) measurements on the graphene part of the TI/G heterostructure device.** The non-local resistance $R_{NL} = V_{NL}/I$ as a function of in-plane magnetic field measured at I=5 µA (a) and I=-5 µA (b). The green arrows represent the directions of magnetization in the two Co ("3" & "2") contacts. The dc bias current is applied between the Co ("2") and Au ("1") electrodes, and the non-local voltage ($V_{NL}=V_3-V_5$) is measured between the Co ("3") and Au ("5") electrode. All the measurements are performed at T=1.6 K.

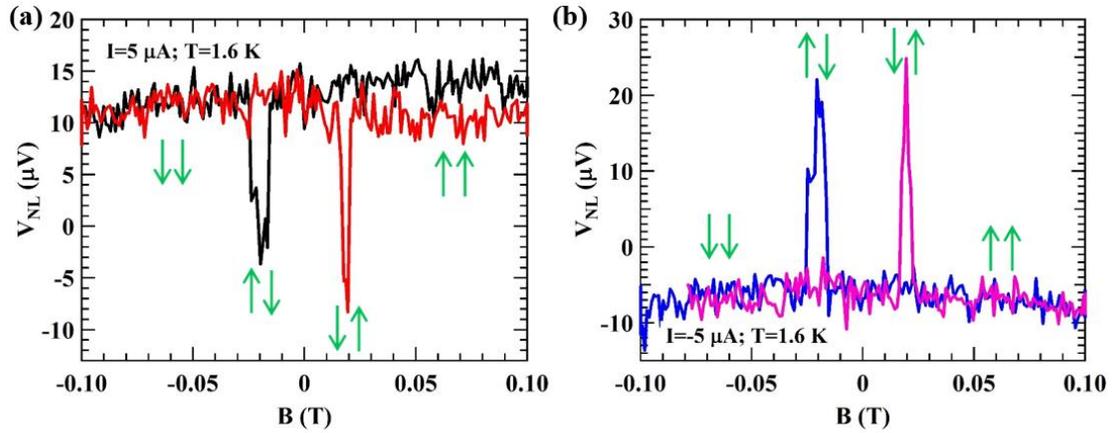

**Figure S4. Non-local spin valve measurements on the graphene part of the TI/G heterostructure.** A dc bias current is applied between the Co ("3") and Au ("5") electrodes, and the non-local voltage $V_{NL}$ is measured between the Co ("2") and Au ("1") electrode. The non-local voltages $V_{NL}$ measured at (a) I=5 µA or (b) I=-5 µA are plotted as a function of in-plane magnetic field. The green arrows represent the directions of magnetization in the Co ("3" and "2") contacts. All the measurements are measured at T=1.6 K.

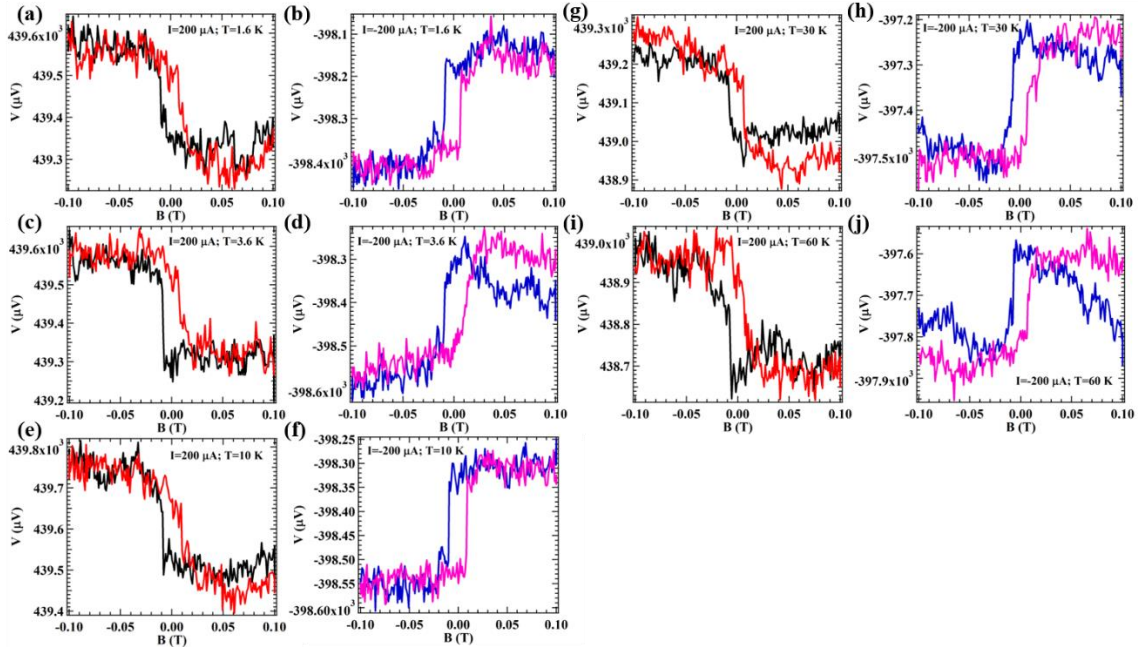

**Figure S5. Complete datasets of the temperature dependence of the measured spin signal shown in Fig. 4e**. (a-j) Voltage measured by the spin detector (Fig. 4a) as a function of the in-plane magnetic field at various temperatures (*T*) ranging from (a,b)1.6 K, (c,d) 3.6 K, (e,f) 10 K, (g,h) 30 K, and 60 K. The applied dc bias currents are 200 μA (a,c,e,g,i) and -200 μA (b,d,f,h,j), respectively.

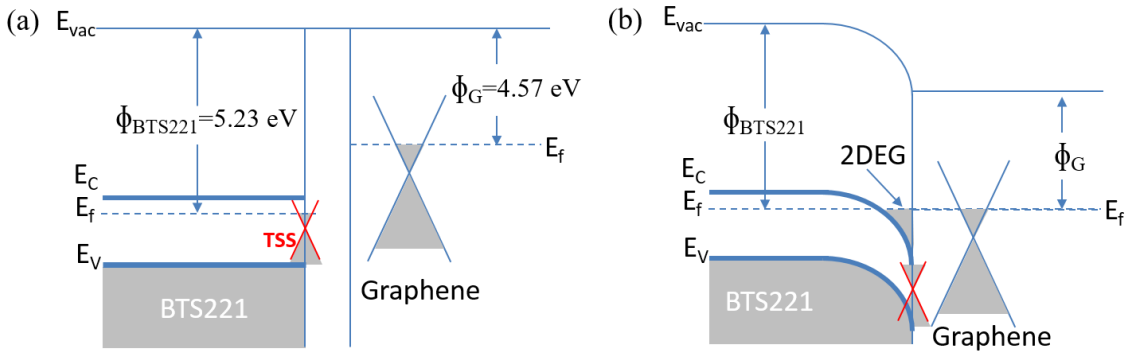

**Figure S6 Band bending and formation of the Rashba 2DEG at the BTS221/graphene interface.** (a-b) Energy band diagram of BTS221 and graphene when they are separated (a) and (b) in contact. A band bending [S1-3] occurs at the BTS221-graphene interface (the bottom surface of TI in our structure) due to the large difference of their work functions as discussed in the main text, forming a Rashba 2DEG derived from the bulk conduction band electrons in the BTS221.